# COMMENTS ON 'CONTROVERSY ON A DISPERSION RELATION FOR MHD WAVES' BY CHANDRA AND KUMTHEKAR

V. S. PANDEY[1] and B. N. DWIVEDI[2]

*(1) Udaipur Solar Observatory, Physical Research laboratory, P. O. Box 198, Dewali, Badi Road, Udaipur- 313001, India.*

*(2) Department of Applied Physics, Institute of Technology, Banaras Hindu University, Varanasi-221005, India.*

Email: pandey_vs@yahoo.com

We comment on the work by Chandra and Kumthekar (2007, henceforth CK) which is questionable. In the derivation of dispersion relation, CK neither invoke the concept of vector space nor do they follow the basic criterion for the elimination of perturbation terms under which the damped magnetoacoustic waves are derived.

**Criteria for deriving the dispersion relation:** For the sake of clarity, we first describe the basic importance of the co-ordinate system under which the motions are described. All the earlier authors (cf., Porter, Klimchuk and Sturrock, 1994, henceforth, PKS; Kumar, Kumar and Singh, 2006, henceforth, KKS; Pandey and Dwivedi, 2007, henceforth PD) have chosen a co-ordinate system in which its z-axis lies along the background magnetic field ($B_0$). Propagation vector $\boldsymbol{k} = k_x \hat{x} + k_z \hat{z}$ lies in the x-z plane, while y-axis is normal to both propagation vector and background magnetic field as shown in Figure1. The beauty of choosing this co-ordinate system is that it contains simultaneously the fluctuations for both incompressible and compressible fluids. This provides criteria in which one derives the dispersion relation. If we consider motion along y-axis for which $\nabla \cdot v = 0$, $i.e., k_y \cdot v_y = 0$, it defines the motion for an incompressible fluid and a criterion for deriving the dispersion relation. In order to obtain dispersion relation, we need to eliminate other fluctuating terms in terms of $v_y$. Also, if we consider motion along x-z plane for which $\nabla \cdot v \neq 0$, $i.e., k_x \cdot v_x + k_z \cdot v_z \neq 0$, it defines motion for compressible fluid, and a criterion for deriving the dispersion relation. This simply means that to obtain the dispersion relation, we need to eliminate the other fluctuating terms in terms of $v_x$ and $v_z$. This was precisely the reason why Carbonell, Oliver, and Ballester (2004); and Carbonell et al. (2006) eliminated other fluctuating terms in terms of $v_x$ and $v_z$ to obtain two algebraic equations in terms of $v_x$ and $v_z$. The solution of these equations yield a dispersion relation for damped magnetoacoustic waves.

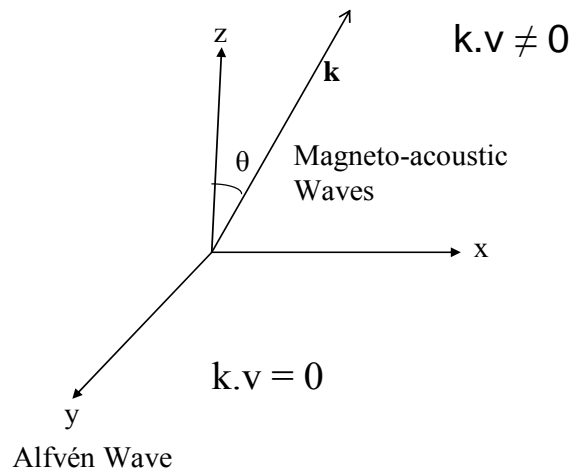

*Figure* 1: A sketch of the co-ordinate system

Although Carbonell, Oliver, and Ballester (2004) have reported a 5-order dispersion relation, they have described their study for a different background configuration, in which they have assumed the background magnetic field along the horizontal axis (i.e., x-axis) instead of vertical axis (i.e., z-axis). Apart from this, it is also noted that Carbonell, Oliver, and Ballester (2004) have not studied the effect of viscosity in their derivation.

The flaws in the CK work are listed below.

1. **Induction equation:** We have taken the same induction equation as given in PKS. However, KKS have taken a different form of this equation. But the linearized form of these set of equations would be the same as also noted by CK. Since dispersion relation is derived from the linearized equations, there should be no difference in the dispersion relation, whether one considers our set of induction equation or that of KKS.

2. **Energy equation:** The claim made by CK for the energy equation is wrong. Substituting continuity equation in PD or PKS energy equation, one will get KKS energy equation, and vice versa. It is just the different form of representation of the energy equation.

3. If we see equations (11) - (19) of PD or equations (24) - (32) of KKS, we get the perturbation terms $\rho_1, p_1, T_1, B_{1x}$ and $B_{1z}$ which are written as a linear combination of $v_{1x}$ and $v_{1z}$. Therefore, one cannot consider $\rho_1, p_1, T_1, B_{1x}$ and $B_{1z}$ as independent variables by the well known concept of vector space. We can only consider $v_{1x}$ and $v_{1z}$ as independent variables.

4. The determinants (20) and (35) proposed by CK is questionable, because it is derived by assuming the dependent variables $\rho_1, p_1, T_1, B_{1x}$ and $B_{1z}$ as an independent variables.

5. Since the foundation of the determinants (20) and (35) are wrong, the common determinant (e.g., section 4, of CK) which is derived after reducing determinants (20) and (35) is also wrong.

6. Since there are two independent variables $v_{1x}$ and $v_{1z}$, therefore all other perturbation terms like $\rho_1, p_1, T_1, B_{1x}$ and $B_{1z}$ are eliminated in terms of $v_{1x}$ and $v_{1z}$, thereby obtaining two algebraic equations for velocity perturbations, corresponding to $2 \times 2$ determinant and solution of this determinant yields a 6-order dispersion relation (cf., PD, 2007). Similar approach has also been considered by PKS as well as Carbonell, Oliver, and Ballester (2004).

7. Braginskii (1965) has also considered a similar approach for deriving the dispersion relation for ideal plasma (cf., equations, 8.22a, 8.22b, 8.23 of Braginskii, 1965).

8. The common determinant (section 4) proposed by CK is a $3 \times 3$ determinant which means that there are three independent variables instead of two, which is wrong. Since damped magnetoacoustic waves are defined in the x-z plane as illustrated in Braginskii (1965) and PKS (1994) by assuming a co-ordinate system as discussed above (cf., Figure 1). The complete motion is defined by two independent variables $v_{1x}$ and $v_{1z}$. Thus an additional independent variable, considered by CK and also by KKS is a wrong approach.

9. The paper by Dwivedi and Pandey (2003) has been retracted by the authors (cf., author comments on the NASA-ADS website).

## Conclusion

In conclusion, we suggest that one should take care of the co-ordinate system under which the motion is defined to derive a dispersion relation. The number of independent variables is defined by the dimensionality of space as per the concept of vector space. Thus a 6-order dispersion relation reported by PD is entirely correct.